\documentclass[times,10pt,twocolumn]{article}
\usepackage{latex8}
\usepackage{times}

\usepackage{graphicx}% Include figure files
\usepackage{dcolumn}% Align table columns on decimal point
\usepackage{bm}% bold math
\usepackage{cite}

\usepackage{amssymb,latexsym}
\usepackage{amsmath}
\usepackage{setspace}
\usepackage{subfigure}
\usepackage{subfigmat}
\usepackage{algorithmic}
\usepackage{algorithm}
\newtheorem{qTheorem}{Theorem}
\newtheorem{definition}{Definition}

\begin{document}

\title{A General Framework for Scalability and Performance Analysis of DHT Routing Systems}
\author{Joseph~S.~Kong, Jesse~S.~A.~Bridgewater and Vwani~P.~Roychowdhury\\
Department of Electrical Engineering\\ University of California, Los Angeles\\
\{jskong, jsab, vwani\}@ee.ucla.edu } 

\maketitle

\begin{abstract}
In recent years, many DHT-based P2P systems
have been proposed, analyzed, and certain
deployments have reached a global scale with nearly one million nodes.
One is thus faced with the question of which particular DHT system to choose,
and whether some are inherently more robust and scalable. 
 Toward developing such a comparative framework, we present
the reachable component method (RCM) for analyzing the performance of different DHT
routing systems subject to {\em random failures}.  We apply RCM to five DHT systems and obtain analytical
expressions that characterize their \emph{routability} as a
continuous function of system size and node failure probability.  An important
consequence is that in the large-network limit, the routability of certain DHT systems go to zero for 
 {\em any} non-zero probability of node failure. These
DHT routing algorithms are therefore \emph{unscalable}, while some others, including
Kademlia, which powers the popular eDonkey P2P system, are found to be \emph{scalable}. 
\end{abstract}
 
\section{ Introduction}\label{sec:intro}
Developing scalable and fault tolerant systems to leverage and utilize the shared resources of distributed computers 
has been an important research topic since the dawn of computer networking.  In recent years, the popularity and 
wide deployment of peer-to-peer (P2P) systems has inspired the development of distributed hash tables (DHTs).  DHTs
typically offer scalable $O(\log n)$ routing latency and efficient lookup interface.  According to a recent 
study \cite{Parker:cachelogic}, the DHT based file-sharing network eDonkey is emerging as 
one of the largest P2P systems with millions of users and accounting 
for the largest fraction of P2P traffic, while P2P traffic currently accounts for 60\% of the 
total Internet bandwidth.  Given the transient nature of P2P users, analyzing and understanding the 
robustness of DHT routing algorithms in the asymptotic system size limit under unreliable environments become essential.  

In the past few years, there has been a growing number of newly proposed DHT routing algorithms.  
However, in the DHT routing literature, there have been few papers that provide a 
general analytical framework to compare across the myriad routing algorithms.  In this paper, 
we develop a method to analyze the performance and scalability of different DHT routing systems 
under random failures of nodes. We would like to emphasize that we intend to analyze the 
performance of the \emph{basic} routing geometry and protocol.  In a real system implementation, there is 
no doubt that a system designer have many optional features, such as additional sequential neighbors,
to provide improved fault tolerance.  Nevertheless, the analysis of the basic routing geometry
will give us more insights and good guidelines to compare among systems.  
%It is crucial to analyze and characterize routing performance 
%under random failure, since the Internet and the end-users of DHTs are extremely transient. 

In this paper, we investigate the routing performance of five DHT systems with uniform 
node failure probability $q$. Such a failure model, also 
known as the \emph{static resilience} model\footnote{The term \emph{static} refers to the assumption that 
a node's routing table remains unchanged after accounting for neighbor failures.}, is assumed in 
the simulation study done by Gummadi et al. \cite{Gummadi:impact}. 
A static failure model is well suited for analyzing performance in the shorter time scale. In a DHT, very fast detection 
of faults is generally possible through means such as TCP timeouts or keep-alive messages, but establishing new 
connections to replace the faulty nodes is more time and resource consuming. The applicability of the results 
derived from this static model to dynamic situations, such as churn, is currently under study. %however, one 
%would expect the short-term performance characteristics, when rewiring and recovery routines are still in 
%progress, to be modeled well by the static resilience model. 

Intuitively, as the node failure probability $q$ increases, the routing performance of the system will worsen.  
A quantitative metric, called \emph{routability} is 
needed to characterize the routing performance of a DHT system under random failure:
\begin{definition} 
The \emph{routability} of a DHT routing system is the expected number of routable pairs 
of nodes divided by the expected number of possible pairs among the surviving nodes.  
In other words, it is the fraction of survived routing paths in the system.  In general,
routability is a function of the node failure probability $q$ and system size $N$.
\label{def:routability}  
\end{definition}

As the DHT-based eDonkey is reaching global scale, it is important to study how 
DHT systems perform as the number of nodes reaches millions or even billions. In fact, we know 
from site percolation theory\cite{Stauffer:intro_perc}, that if $q>(1-p_c)$, where $p_c$ is called the 
percolation threshold of the underlying network, then the network will get fragmented into very small-size 
connected components and for large enough network size. As a result, the routability of the network will 
approach zero for such failure probability due to the lack of connectivity.  
%Such percolation results, however, hold only for infinite-size networks, and there is no general result 
%on the size of the largest connected components as a continuous function of failure probability, $q$, and network size $N$. 
However, because of how messages get routed as specified by the underlying routing protocol, all pairs 
belonging to the same connected component need not be reachable under failure.

In general, the size of the connected 
components do not directly give us the routability of the subnetworks. Hence, one needs to develop a framework 
different from the well-known framework of percolation. As a result, this 
work investigates DHT routability under the random failure model for both finite system sizes 
and the infinite limit. We will define the \emph{scalability} of a routing system as follows: 
\begin{definition}
A DHT routing system is said to be \emph{scalable} 
if and only if its routability converges
to a nonzero value as the system size goes to infinity for a nonzero failure probability $q$.  
Mathematically, it is defined as follows:
\begin{equation*}
 \lim_{N \rightarrow \infty} r(N,q) > 0 \ \ for \ 0 < q < 1 - p_c
\end{equation*}
where $r(N,q)$ denotes the routability of the system as a function of system size $N$ and 
failure probability $q$.  Similarly, the system is said to be \emph{unscalable} if and only if 
its routability converges to zero as the system size goes to infinity for a nonzero failure 
probability $q$:
\begin{equation*}
 \lim_{N \rightarrow \infty} r(N,q) = 0 \ \ for \ 0 < q < 1 - p_c
\end{equation*}
\label{def:scalability}
\end{definition}
We want to emphasize that in a real implementation, there are many system parameters 
that the system designer can specify, such as the number of near neighbors or 
sequential neighbors.  As a result, the designer can always 
add enough sequential neighbors to achieve an acceptable routability under reasonable 
node failure probability for a maximum network size
that exceeds the expected number of nodes that will participate in the system. 
The scalability definition 
is provided for examining the \emph{theoretical} asymptotic behavior of DHT routing 
systems, not for claiming a DHT system is unsuitable for any large-scale deployment.    

Having specified the definition of the key metrics, we will present the reachable component method (RCM), a
simple yet effective method for analyzing DHT routing performance under random failure. 
We apply the RCM method to analyze the basic routing algorithms used in the following five DHT systems: 
Symphony \cite{Manku:symphony}, Kademlia \cite{May:kademlia}, Chord \cite{Stoica:chord}, CAN \cite{Ratnasamy:CAN} and 
Plaxton routing based systems \cite{Plaxton:routing}.  For all algorithms except Chord routing, we 
derive the analytical expression for each algorithm's routability under random failure, 
while an analytical expression for a tight lower bound is obtained for Chord routing.
In fact, our analytical results match the simulation results carried out in \cite{Gummadi:impact}, 
where different DHT systems were simulated and the percentage of failed paths (i.e., 1-routability)
 was estimated for $N=2^{16}$, as illustrated in Fig. \ref{fig:compare_fig}.  
In addition, we also derive the asymptotic performance of the routing algorithm under failure 
as the system scales.  

One interesting finding of this paper is that under random failure, 
the basic DHT routing systems can be classified into two classes: \emph{scalable} and \emph{unscalable}. 
For example, the XOR routing scheme of Kademlia is found to be \emph{scalable}, since the routability 
of the system under nonzero probability of failure converges very fast to a positive limit even 
as the size of the system tends to infinity.  This is consistent with the observation that the
Kademlia-based popular P2P network eDonkey is able to scale to millions of nodes.  
In contrast, as the system scales, the routability 
of Symphony's routing scheme is found to quickly converge to zero for any failure probability 
greater than zero.  Thus, the basic routing system for Symphony is found to be \emph{unscalable}.
However, as briefly discussed above in this section, a system designer for Symphony can specify 
enough near neighbors to guarantee an acceptable routability in the system for a maximum network 
size and a reasonable failure probability $q$.

The rest of this paper is organized as follows.  In section \ref{sec:related}, we discuss previous 
work on the fault tolerance of P2P routing systems.  
In section \ref{sec:overview}, we will give an overview of the DHT routing systems that we 
intend to analyze.  
In section \ref{sec:rcm}, we present the \emph{reachable component method} (RCM) and apply the RCM 
method on several DHT systems.  
In section \ref{sec:scalability}, we examine the scalability of DHT routing systems. 
In section \ref{sec:conclusion}, we give our concluding remarks.  
 
\section{Related Work}\label{sec:related}
The study of robustness in routing networks has grown in the past few years with researchers simulating 
failure conditions in DHT-based systems. Gummadi et al.\cite{Gummadi:impact} showed through simulation 
results that the routing geometry of each system has a large effect on the network's static resilience 
to random failures.  
\iffalse Static resilience represents the performance of the system subject to failures when no measures are taken to compensate for connectivity loss. This is not only a realistic model for many systems, but it also represents a baseline of performance against which different recovery algorithms can be compared. In Section \ref{sec:intro} we note that detecting a failed connection can be very fast, thus when a node's neighbor becomes unavailable the node will not attempt to incorrectly route packets through that neighbor for long.  However, while detection of failure is fast, it will not generally be sufficiently fast to create new connections to compensate for the lost connectivity. The existence of these two time-scales  makes the static resilience failure model important and relevant. We do not consider recovery algorithms and other dynamics here but will instead focus on the static random failure model. \fi
In addition, there have been research work done in the area of analyzing and simulating 
dynamic failure conditions (i.e. churn) in DHT systems \cite{Liben:analysis,Li: comparing,Krish:chord}.

Theory work has been done to predict the performance of DHT systems
under a static failure model.  The two main approaches thus far have been graph theoretic
methods\cite{Angel:routing,Loguinov:graph,Lam:evaluation} and Markov
processes\cite{Wang:markov}. Most analytical work to date has dealt with one
or two routing algorithms to which their respective methods are well-suited
but have not provided comparisons across a large fraction of the DHT
algorithms. Angel et al. \cite{Angel:routing} use percolation theory to place
tight bounds on the critical failure probability that can support efficient routing on
both hypercube and $d$-dimensional mesh topologies.  By efficient they mean
that it is possible to route between two nodes with time complexity on the
order of the network diameter.  While this method predicts the point at which the network
becomes virtually unusable, it does not allow the detailed
characterization of routability as a function of the failure probability.  
\iffalse Wang et al.\cite{Wang:markov}
model CAN with small-world extensions using Markov-chain methods. This method
is straightforward and produces detailed performance predictions. However
the method's usefulness is limited somewhat by the fact that the network 
connectivity structure is represented as a matrix and consequently the level 
of detail in the model approaches that of a routing simulation. Because of
this approach, a fairly-complex numerical computation must be done for each
system size and node failure probability to yield the routability. \fi
In contrast, the reachable component method (RCM) method exploits the geometries of DHT routing 
networks and leads to simple analytical results that predict routing performance 
for arbitrary network sizes and failure probabilities.

\section{Overview of DHT Routing Protocols}\label{sec:overview}
%We will fill this section in.  Basically, we need to re-word section 2.3 of the Gummadi et al. 
%Impact of DHT Routing Geometry paper, which is included in the tarball for your reference: 
%papers/p381-gummadi.pdf .  
We will first review the five DHT routing algorithms that we intend to analyze.
An excellent discussion of the geometric interpretation of these routing algorithms 
(except for Symphony) is provided by Gummadi et al.\cite{Gummadi:impact} and we use the same terms
for the geometric interpretations of DHT routing systems in this paper (e.g. hypercube 
and ring geometry for CAN and Chord routing systems, respectively).
By following the algorithm descriptions in \cite{Gummadi:impact} as well as the 
descriptions in this section, one can construct Markov chain models (e.g. Fig. \ref{fig:two_mc}) 
for the DHT routing algorithms.  The application of the Markov chain models will be discussed in 
section \ref{ssec:rcm_des} and \ref{ssec:hyper_example}.
%These Markov chain models can then be used to obtain expressions for
%$p(h,q)$ for each algorithm. 

In addition, we will use the notation of \emph{phases} as used in 
\cite{Kleinberg:algorithmic}: we say that the routing process has reached phase 
$j$ if the numeric distance (used in Chord and Symphony) or the XOR distance (used in 
Kademlia) from the current message holder to the target is 
between $2^j$ and $2^{j+1}$. In addition, we will 
use binary strings as identifiers although any other base besides 2 can be used.  
Finally, for those systems that require resolving node identifier bits 
\emph{in order}, we use the convention of correcting bits from left to right. 
\subsection{Tree (Plaxton)}
Each node in a tree-based routing geometry has $\log N$ neighbors, with the 
$i$th neighbor matching the first $i-1$ bits and differ on the $i$th bit.
When a source node $S$, wishes to route to a destination, $D$, the routing can 
only be successful if one of the neighbors of $S$ , denoted $Z$,
shares a prefix with $D$ and has the highest-order differing bit. 
Each successful step in the routing results in the highest-order bit being
corrected until no bits differ.
 
The routing Markov chain (Fig. \ref{fig:plaxton_mc}) for the tree geometry can easily be generated by
examining the possible failure conditions during routing. At each step in the routing process, the 
neighbor that will correct the leftmost bit must be present in order for the message to be routed.
Otherwise, the message is dropped and routing fails. 
\subsection{Hypercube (CAN)}
In the hypercube geometry, each node's identifier is a binary string
representing its position in the $d$-dimensional space. The distance between
nodes is simply the Hamming distance of the two addresses. The number of possible paths that can correct a
bit is reduced by 1 with each successful step in the route.
This fact makes the creation of the hypercube routing Markov
chain (Fig. \ref{fig:hypercube_mc}) straightforward.
\subsection{XOR (Kademlia)}
In XOR routing \cite{May:kademlia}, the distance between two nodes is the numeric 
value of the XOR of their node identifiers.  Each node keeps $\log(N)$ connections, 
with the $i$th neighbor chosen uniformly at random from an XOR distance in the range of 
$[2^{d-i},2^{d-i+1}]$ away. Messages are 
delivered by routing greedily in the XOR distance at each hop.  Moreover, it is a simple exercise to show 
that choosing a neighbor at an XOR distance of $[2^{d-i},2^{d-i+1}]$ away is equivalent to choosing a 
neighbor by matching the first (i-1) bits of one's identifier, flipping the $i$th bit, 
and choose random bits for the rest of the bits. 

Effectively, this construction is equivalent to the Plaxton-tree routing geometry. 
As a result, when there is no failures, the XOR routing protocol resolves node 
identifier bits from left to right as in the Plaxton-tree geometry.  However, 
when the system experiences node failures, nodes have the option to route 
messages to neighbors that resolve lower order bits when the neighbor that would 
resolve the highest order bit is not available.  Note that resolving lower order bits
will also make progress in terms of decreasing the XOR distance to destination.  
Nonetheless, the progress made by resolving lower order bits is not necessarily 
preserved in future hops or phases (see Fig. \ref{fig:xor_routing}). 

For example, at the start of the routing process, one phase is advanced if 
the neighbor correcting the leftmost bit exists.  Otherwise, the routing process 
can correct one of the lower order bits. However, if all of the neighbors that would
resolve bits have failed, the routing process fails.  A Markov chain model for the routing 
process is illustrated in Fig. \ref{fig:xor_mc}.   

\subsection{Ring (Chord)}
In Chord \cite{Stoica:chord}, nodes are placed in numerical order around a ring.  
Each node with identifier $a$ maintains $\log(N)$ connections 
or fingers, with each finger at a distance $[2^{d-i},2^{d-i+1}]$ away (the randomized 
version of Chord is discussed here). Routing can be done greedily on the ring.  
When the system experiences failure, each node will continue to route a message 
to the neighbor closest to destination (i.e. in a greedy manner).  
A Markov chain model for the routing process is illustrated in Fig. \ref{fig:chord_mc}.   

\subsection{Small-World (Symphony)}
Small-world routing networks in the $1$-dimensional case have a ring-like
address space where each node is connected to a constant number of its nearest
neighbors and a constant number of shortcuts that have a $1/d$ distance
distribution ($d$ is the ring-distance between the end-points of the shortcut).
Each node maintains a constant number of neighbors and uses greedy routing.
Due to the distance distribution it will take an average of $O(\log N)$ hops
before routing halves the distance to a target node, therefore requiring
$O(\log N)$ such phases to reach a target node for a total expected latency of
$O(\log^2 N)$. 

When the system experiences node failures, some of the shortcuts will be
unavailable and the route will have to take "suboptimal" hops.
The small-world Markov chain model is fundamentally
different from the ones for XOR routing (Fig. \ref{fig:xor_mc}) and ring routing (Fig. \ref{fig:chord_mc}). 
A routing phase is completed if any of the node's 
shortcuts connects to the desired phase. This happens with probability $\frac{k_s}{d}$ where
$k_s$ denotes the number of shortcuts that each node maintains.  Alternatively, the 
routing fails if all of the node's near neighbor and shortcut connections fail, which 
happens with probability $q^{k_n+k_s}$.
If neither of the above happens then the route takes a suboptimal hop, which happens with probability 
$1-\frac{k_s}{d}-q^{k_n+k_s}$.  

\section{Reachable Component Method and its Applications}\label{sec:rcm}
\begin{figure*}
%\centering
\begin{tabular}{lll}

\begin{minipage}[b]{1.5in}
\includegraphics[width=1.5in]{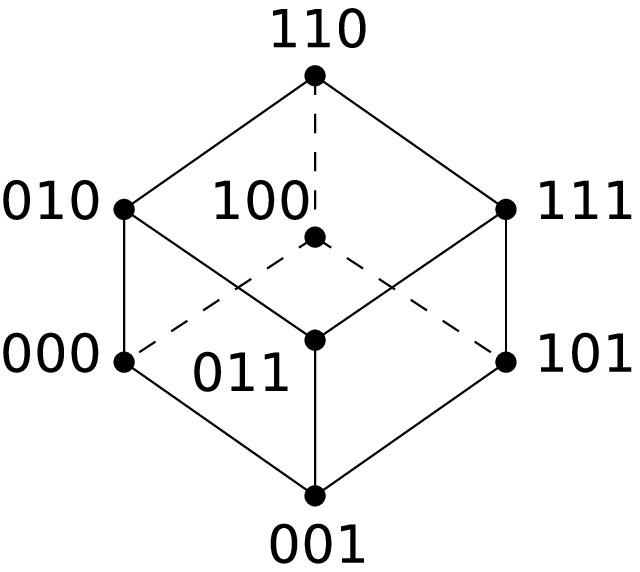}
\caption{\small Here we illustrate the reachable component method using an 8-nodes hypercube.}
\label{fig:merge_1}
\end{minipage}

& \begin{minipage}[b]{1.9in}
\includegraphics[width=1.5in]{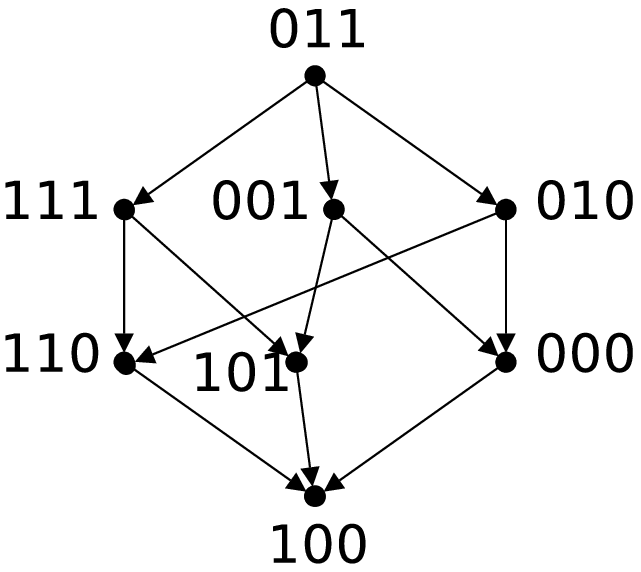}
\caption{\small We select node 011 to be the root of the routing graph. The symmetry of the system 
means that each node will be the root of a routing graph with identical structure.}
\label{fig:merge_2}
\end{minipage}
&
\begin{minipage}[b]{3.1in}
\begin{minipage}[b]{3.1in}
\includegraphics[width=1.5in]{figs/hc_routing}\nobreak
\begin{array}[b]{c c c c}
$$h$$ & $$n(h)$$ & $$\mbox{Pr}(S_h,S_{h+1})$$ \\
\hline\\
$$1$$ & $$\binom{3}{1}$$ & $$1-q^3$$ 
\\
\\
$$2$$ & $$\binom{3}{2}$$ & $$1-q^2$$ 
\\
\\
$$3$$ & $$\binom{3}{3}$$ & $$1-q$$ 
\\
\\
\end{array}
\end{minipage}
\caption{\small For illustration purpose, we examine how 011 routes a message to 100.  
Note that three choices exist for the first hop, 2 choices exist for the second hop and only 
one choice left for the last hop.  For this example, $p(h,q)$ is: 
$ p(3,q) = (1-q^3)(1-q^2)(1-q) $.}
\label{fig:merge_3}
\end{minipage}

\end{tabular}
\end{figure*}

\subsection{Method Description} \label{ssec:rcm_des}
We now describe the steps of the \emph{reachable component method} (RCM) in calculating the 
routability of a DHT routing system under random failure.  Before we delve into the description, 
let us first clarify several concepts and notations on DHT routing.  First, we allow all DHTs 
to fully populate their identifier spaces (i.e. node identifier length $d = \log_b N$). 
%Second, we assume a \emph{static resilience} model as in \cite{Gummadi:impact}, which means that no recovery algorithm
%is employed to repair failed connections, but entries of the routing tables of the failed node's 
%neighbors are modified accordingly. 
Second, when a DHT is not in its perfect topological state, it can be the case that a pair of nodes are 
in the same connected component but these two nodes cannot route between each other.  Thus, the reachable 
component of node $i$ is the set of nodes that node $i$ can 
route to under the given routing algorithm.  Note that the reachable component of node $i$ is a subset of 
the connected component containing node $i$. Third, we assume that no "back-tracking" is allowed 
(i.e. when a node cannot forward a message further, the node is not allowed to return the message 
back to the node from whom the message was received).  

RCM is fairly simple in concept and involves the following five steps: 
\begin{enumerate}
\item Pick a random node, node $i$, from the system and denote it as the \emph{root node}.
Construct the root node's routing topology from 
the routing algorithm of the system (i.e. the topology by which the root node routes to all other 
nodes in the system).
\item Obtain the distribution of the distances (in hops or in phases) between the root node and all other nodes (denoted 
as $n(h)$); in other words, for each integer $h$, calculate the number of nodes at distance $h$ hops 
from the root node. Note that the meaning of \emph{hops} or \emph{phases} will be clear from the context.
\item Compute the probability of success, $p(h,q)$, for routing to a node $h$ hops away from the root node under 
a uniform node failure probability, $q$.  
\item Compute the expected size of the \emph{reachable component} from the root node by first 
calculating the expected number of reachable nodes at distance $h$ hops away (which is simply given by 
$n(h)*p(h,q)$).  Now, we sum over all possible number of hops to obtain the expected size of the reachable
component.  
\item By inspection, the expected number of routable pairs in the system is given by 
summing all surviving nodes' expected reachable component sizes.  Then, dividing the expected number of routable 
pairs by the number of possible node pairs among all surviving nodes produces the 
routability of the system under uniform node failure probability $q$.    
\end{enumerate}

The formula for computing the expected size of the reachable component, $E[S_i]$, described in step 4 
is derived as follows:  
\begin{eqnarray}
E[S_i] &=& E[\displaystyle\sum_{\substack{j=1\\j\neq i}}^{N} Y_j] = \displaystyle\sum_{\substack{j=1\\j\neq i}}^{N} E[Y_j] 
= \displaystyle\sum_{h=1}^{d} n(h)p(h,q) \nonumber
\end{eqnarray}
\noindent where $Y_j$ is Bernoulli random variable for denoting reaching node $j$, 
and $d$ is the node identifier length.
%\noindent where the last equality follows from the observation that the probability of reaching 
%node $j$ only depends on the distance in hops to $j$.  

Since nodes in the system are removed with probability $q$, there are $(1-q)N$ or $pN$ nodes that survive on average. 
In step 5, the formula for calculating the routability, $r$, of the system under uniform
failure probability $q$ is given as follows:  
\begin{eqnarray}
r &=& \frac{M_{rp}}{M_p} = \frac{E\left[\displaystyle\sum_{\small i=1}^{\small pN} S_i\right]}{2\binom{\lfloor pN \rfloor)}{2}} \approx \frac{\displaystyle\sum_{i=1}^{pN} E[S_i]}{pN(pN-1)}\nonumber \\
  &=& \frac{ E[S]}{(pN-1)} \label{eq:routability}
\end{eqnarray}
\noindent where $M_{rp}$ denotes the expected number of \emph{routable pairs} among surviving nodes, 
and $M_p$ is the expected number of all \emph{possible pairs} among surviving nodes.
Note that the last equality follows from the observation that DHTs investigated in this paper 
have symmetric nodes.  Therefore, the routing topology of each node is statistically identical to 
each other.  Thus, all $S_i$'s are identically distributed for all $i$'s: $E[S] = E[S_i] \ \forall i$.

\subsection{Using the Hypercube Geometry as an Example}\label{ssec:hyper_example}
\begin{figure*}[tbh]
\begin{subfigmatrix}{2}
\subfigure[\small Tree]{
\label{fig:plaxton_mc}
\centering
\includegraphics[width=2.8in]{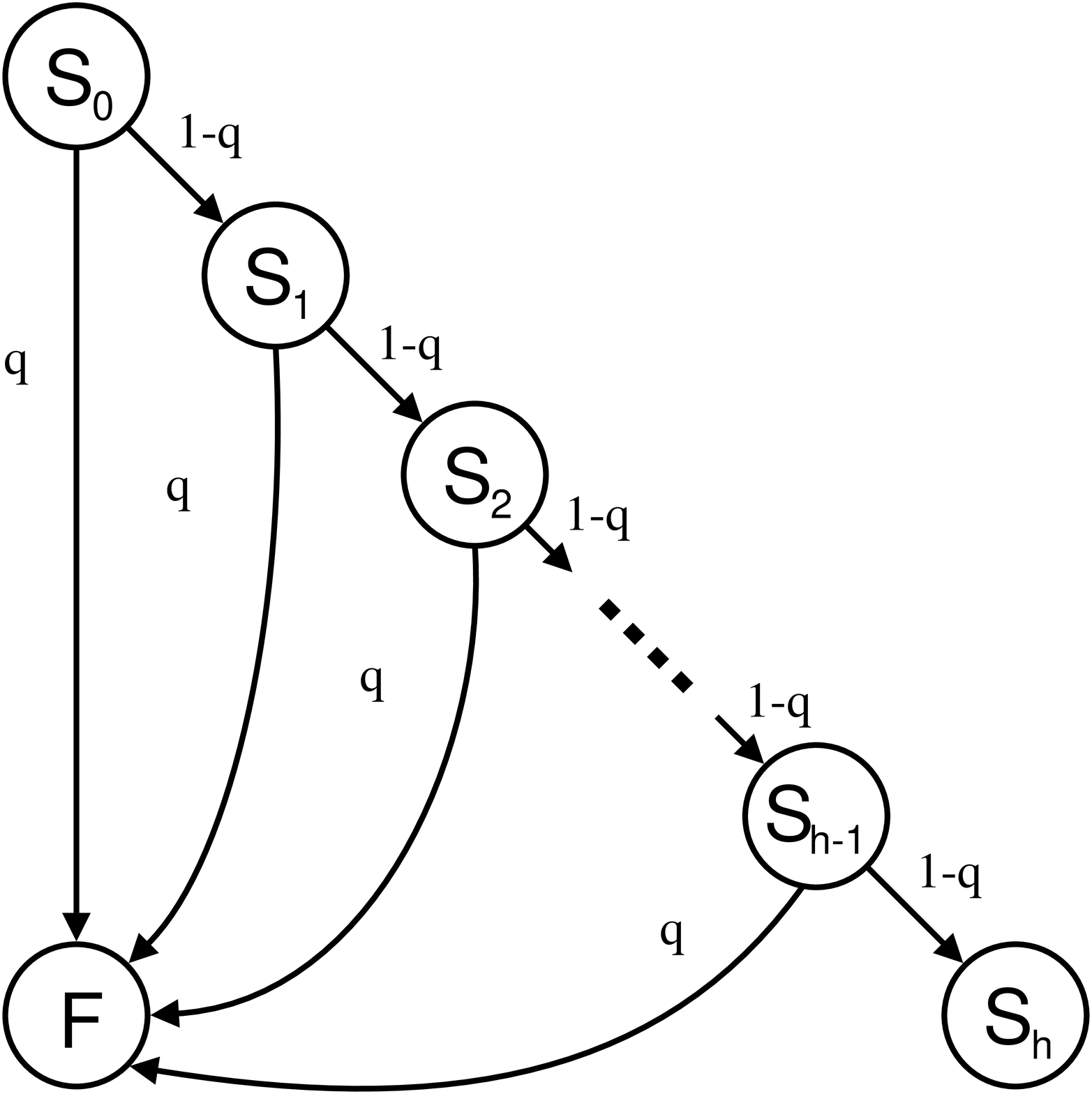}
}
\subfigure[\small Hypercube]{
\label{fig:hypercube_mc}
\centering
\includegraphics[width=2.8in]{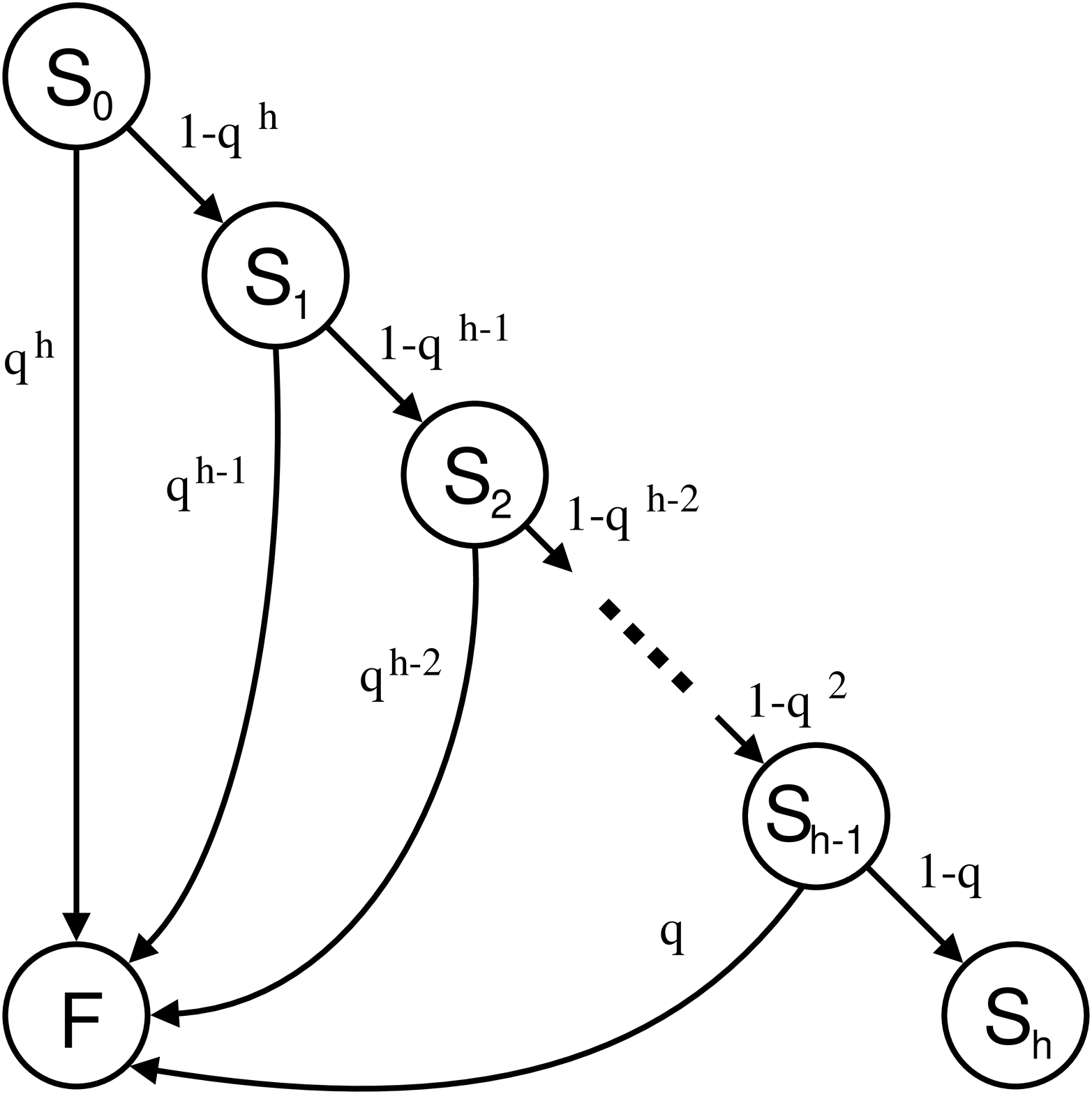}
}
\end{subfigmatrix}
\caption{\small The above diagrams illustrate the Markov chain model of the routing process to a target at distance 
$h$ hops from the root 
node. Note that there are only two absorbing states in these Markov chains: the failure state (denoted 
by $F$) and the success state (denoted by $S_h$).  (a) Markov chain model for tree routing: The $S_i's$ 
represent the states that correspond to number of corrected \emph{ordered} bits. At each $S_i$, 
the neighbor that will correct the leftmost bit must be present in order for the message to be routed.
Otherwise, the message is dropped and routing fails. Thus, the transition probability from $S_i$ to 
$S_{i+1}$ is $1-q$, while the transition probabilities to the failure state is $q$. 
(b) Markov chain model for hypercube routing: Here, the $S_i's$ represent the states that correspond to number 
of corrected bits in any order. The transition probabilities are obtained by noting that at state 
$S_i$, there are $h-i$ neighbors to route the message to. }
\label{fig:two_mc}
\end{figure*}

A simple application of the RCM method is illustrated for the CAN hypercubic routing system in 
Fig. \ref{fig:merge_1}-\ref{fig:merge_3}.  The RCM steps involved are as follows: 

\noindent \textbf{Step 1.} As reviewed in section \ref{sec:overview}, in a hypercube routing 
geometry \cite{Ratnasamy:CAN}, the distance (in hops) between two nodes is their Hamming distance.  
Routing is greedy by correcting bits in any order for each hop. \\
\textbf{Step 2.} Thus, for any random node $i$ in a hypercube routing system with identifier 
length of $d$ bits, we have the following distance distribution: $n(h) = \binom{d}{h}$.
The justification is immediate: a node at $h$ hops away has a Hamming distance of $h$
bits with node $i$.  Since there are $\binom{d}{h}$ ways to place the $h$ differing 
bits, there are $\binom{d}{h}$ nodes at distance $h$ (see Fig. \ref{fig:merge_2}).  \\
\textbf{Step 3.} The routing process can be modeled as a discrete time Markov chain 
(Fig. \ref{fig:merge_3} and \ref{fig:hypercube_mc}).  The states $S_i's$ of the Markov chain 
correspond to the number 
of corrected bits.  Note that there are only two absorbing states in the Markov chain: the 
failure state $F$ and the success state (i.e. $S_h$).  
Thus, the probability of successfully routing to a target node at distance $h$ hops away is 
given by the probability of transitioning from $S_0$ to $S_h$ in the Markov chain model:
\begin{eqnarray}
p(h,q)&=&\mbox{Pr}(S_0 \rightarrow S_1 \rightarrow ... \rightarrow S_h) \nonumber \\
      &=&\mbox{Pr}(S_0 \rightarrow S_1)\mbox{Pr}(S_1 \rightarrow S_2)...\mbox{Pr}(S_{h-1} \rightarrow S_h) \nonumber \\
      &=&(1-q^h)(1-q^{h-1})...(1-q) \nonumber \\ 
      &=&\displaystyle\prod^{h}_{m=1}(1-q^m) \label{eq:hypercube_eq}
\end{eqnarray}
\textbf{Step 4.} Thus, the expected size of the reachable component is given as:
\begin{equation*}
E[S] = \displaystyle\sum_{h=1}^{d} n(h)p(h,q) = \displaystyle\sum_{h=1}^{d} \binom{d}{h}\displaystyle\prod^{h}_{m=1}(1-q^m)
\end{equation*}
\textbf{Step 5.} Using Eq. \ref{eq:routability}, we obtain the analytical expression for routability:
\begin{align}
r &= \frac{\displaystyle\sum_{h=1}^{d} n(h)p(h,q)}{(1-q)2^d -1} \label{eq:routability_eq}\\
  &= \frac{\displaystyle\sum_{h=1}^{d} \binom{d}{h}\displaystyle\prod^{h}_{m=1}(1-q^m)}{(1-q)2^d -1}
\end{align}

\begin{figure*}[tbh]
\subfigure[\small XOR Routing under Failure]{
\centering
\includegraphics[width=3.00in]{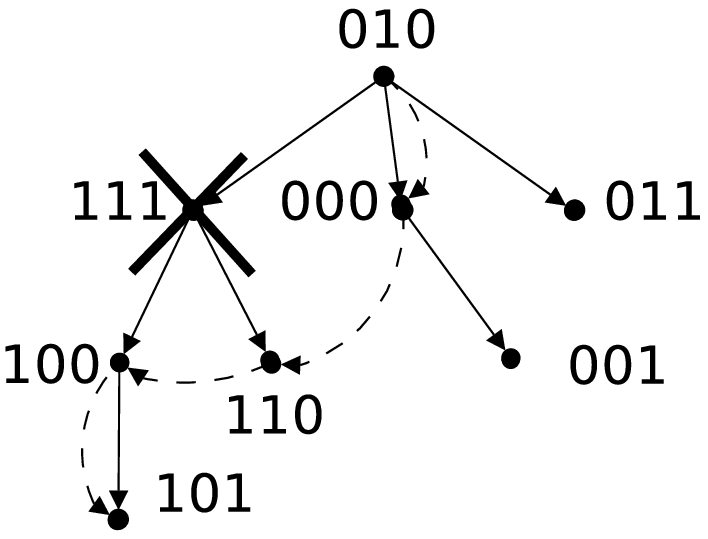}
\label{fig:xor_routing}
}
\subfigure[\small XOR Markov Chain Model]{
\centering
\includegraphics[width=3.70in]{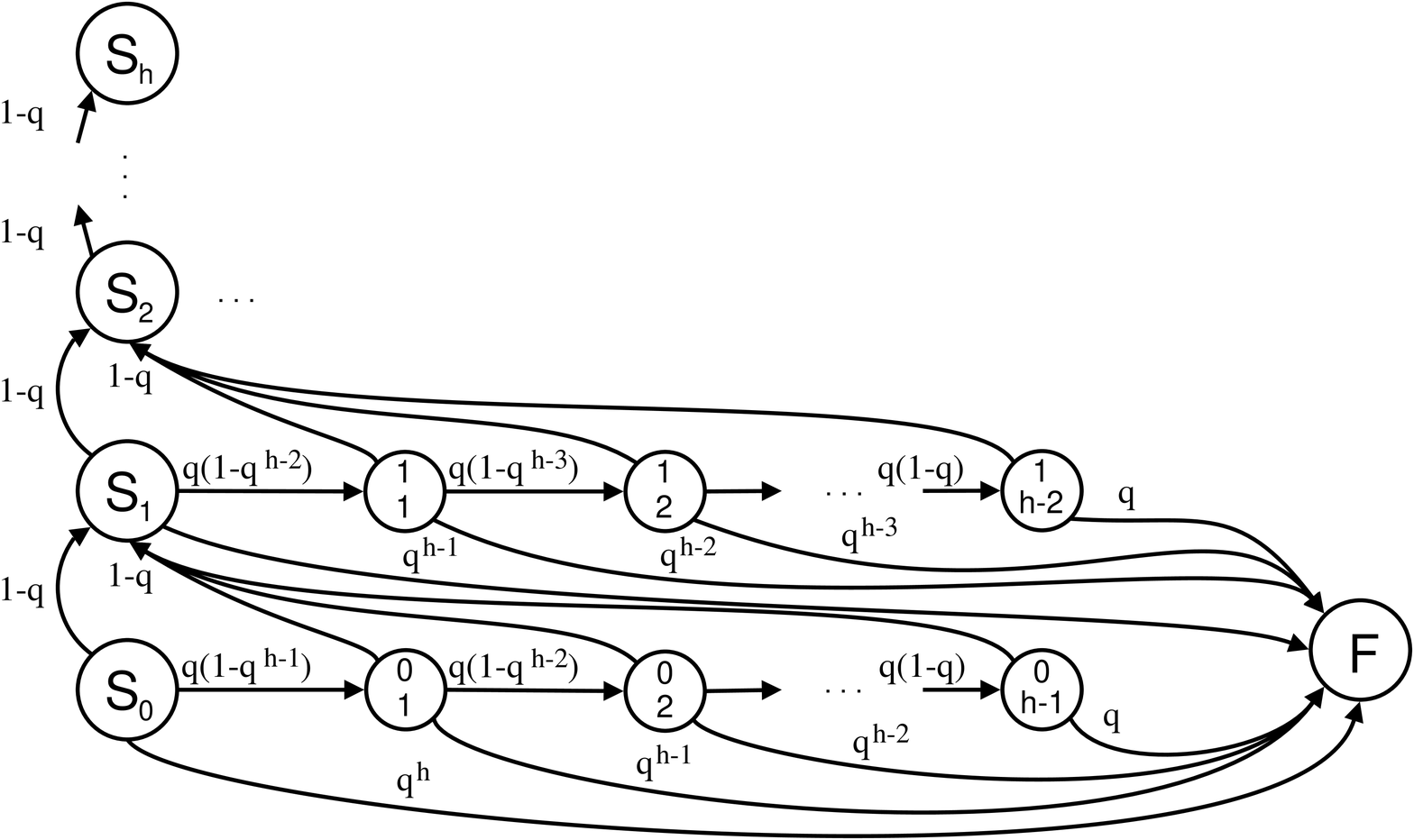}
\label{fig:xor_mc}
}
\caption{\small (a) Illustration of XOR routing under failure: in this simple example, node $010$ tries to route a 
message to node 101.  However, its first neighbor 111 (i.e. the randomly chosen node that flips the first bit 
and chooses random bits for the rest of the identifier bits), has just failed.  As a result, the 
message is routed to node 010's second neighbor, node 000, correcting a lower order bit. Now, node 000's first neighbor, 
node 110, is available, and node 110's second neighbor, node 100, is also available.  
Consequently, the message is routed to the destination node 101, by following the dashed arrows in the diagram.  
(b) Markov chain model for XOR routing: this diagram illustrates routing 
to a target located at $h$ phases in distance, which is equivalent to 
correcting $h$ bits in order (left to right).  The $S_i's$ denote the states that correspond to the number of 
corrected \emph{ordered} bits, which is equivalent to the number of phases advanced. The states $(i,j)$ 
denote a state that corresponds to $j$ suboptimal hops taken after advancing $i$ phases. }
\label{fig:xor_figs}
\end{figure*}

%1. Construct the trellis diagram for computing the probability of routing to a node in $l$ hops.\\
%2. Count the number of nodes at $l$ hops away.\\
%3. Compute $E[S_i]$.\\
%4. Compute routability as the average across all pairs.\\
%5. Emphasize that a Markov Chain is involved.\\

\begin{figure*}[htp]
\begin{subfigmatrix}{2}
\subfigure[]{
\label{fig:analy_vs_sim}
\centering
\includegraphics[width=3.35in]{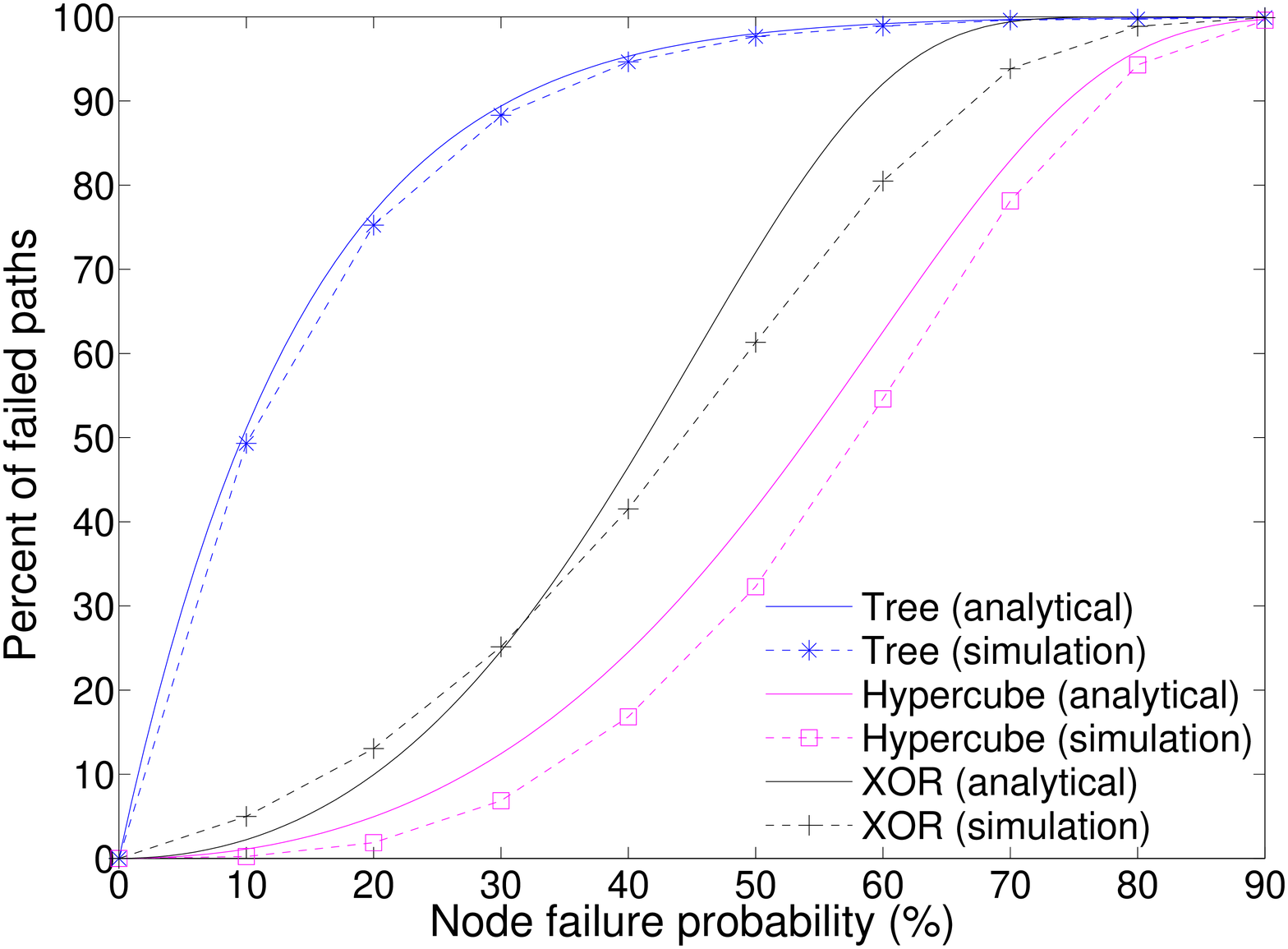}
}
\subfigure[]{
\label{fig:chord_analy_vs_sim}
\centering
\includegraphics[width=3.35in]{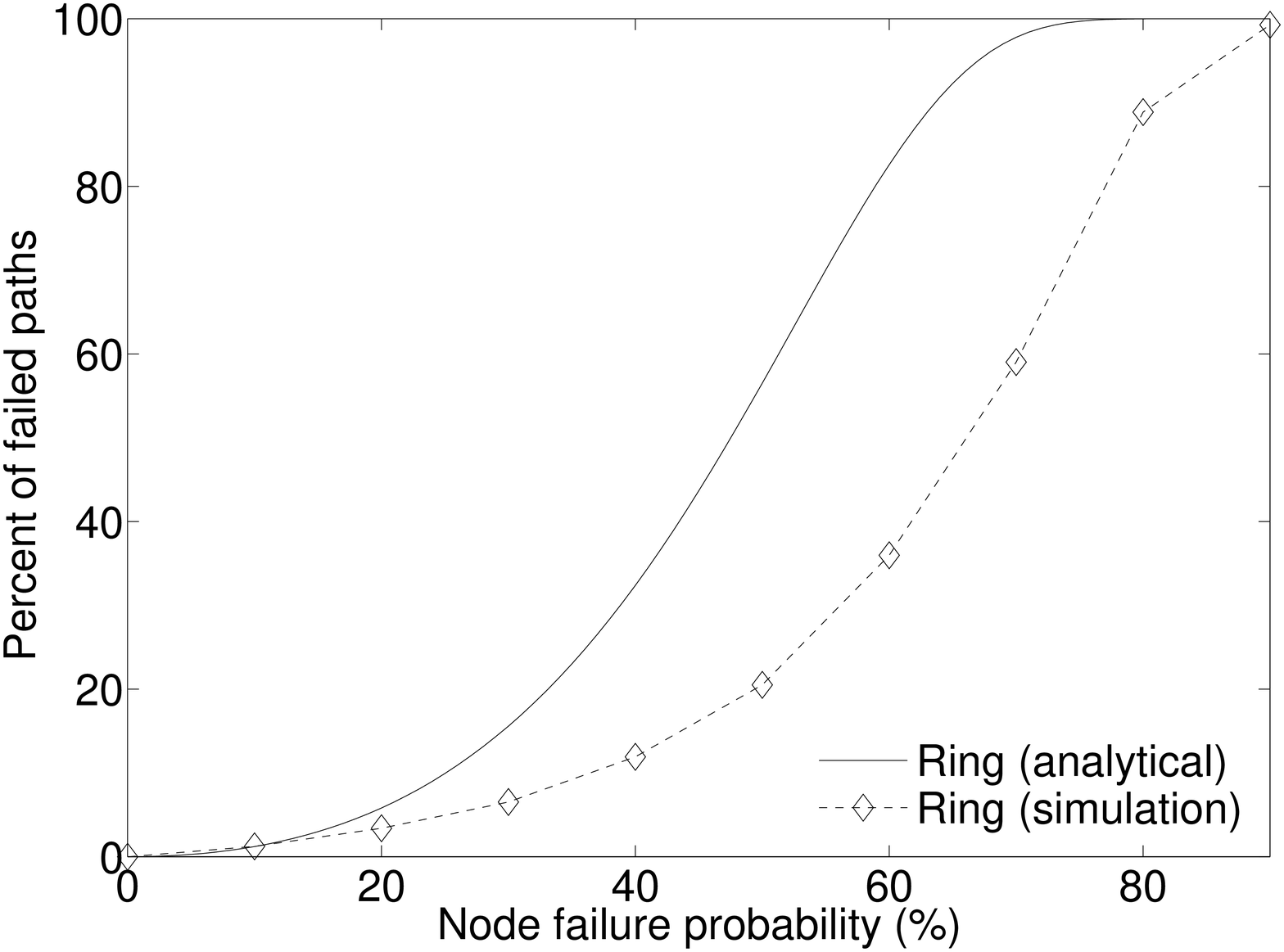}
}
\end{subfigmatrix}
\caption{\small Both plots show the percentage of failed paths (i.e. 1-routability) for 
varying node failure probability and system size of N=$2^{16}$. (a) Analysis vs simulation: The 
simulation data points are reproduced from \cite{Gummadi:impact}. For all three routing 
geometries, the analytical curves show a great fit to the simulation curves.  (b) 
Analysis vs simulation (ring):  For the ring routing algorithm, the discrepancy between the analytical 
and simulation curve is due to the algorithm's property that suboptimal hops contribute 
non-trivially to the routing process.  In effect, the analytical curve provides an upper bound 
for percentage of failed paths.  Note that the analytical curve is very close to simulation 
in the region of practical interest (i.e. for failure probability less than 20\%)}. 
\label{fig:compare_fig}
\end{figure*}

\subsection{Summary of Results for other Routing Geometries}\label{ssec:summary_other}
Using the RCM method, the analytical expressions for the other DHT routing geometries can be similarly derived 
as for the hypercube routing geometry.  In all the derivations, the majority of the work involves finding 
the expression for $p(h,q)$ through Markov chain modeling.  Note that 
the analytical expressions derived in this section are compared with the simulation results 
obtained by Gummadi et al. \cite{Gummadi:impact} in Fig. \ref{fig:analy_vs_sim} and \ref{fig:chord_analy_vs_sim}. 

For ease of exposition, we will use the notation $G(i,j)$, which 
denotes the probability that, starting at state $i$, the Markov chain ever visits state $j$. 
By any of the Markov chain models for the routing protocols, 
we note that $G(S_0,S_1) = 1-Q(h)$, $G(S_1,S_2) = 1-Q(h-1)$, 
and so forth, where the function $Q(m)$ can be thought of as the probability of failure 
at the $m$th phase of the routing process. As a result, all of the DHT systems under 
study have the property that the probability of successfully traveling $h$ hops or phases 
from the root node, $p(h,q)$, is given by the following common form:
\begin{eqnarray}
p(h,q) &=& G(S_0,S_1)G(S_1,S_2)...G(S_{h-1},S_h) \nonumber \\
       &=& \prod^{\small h}_{\small m=1}(1-Q(m)) \label{eq:common_eq}
\end{eqnarray}
Using Eq. \ref{eq:routability_eq}, we see that only the expressions for 
$n(h)$ and $Q(m)$ are needed to compute the routability of the DHT routing system under
investigation.  As a result, we will only provide the $n(h)$ and $Q(m)$ expressions for 
each system for conciseness.  

\subsubsection{Tree} For the tree routing geometry, the routing distance 
distribution, $n(h)$, is $\binom{d}{h}$ by inspection.  Furthermore, 
it is simple to show that $p(h,q) = (1-q)^h$ by examining the Markov chain model 
(see Fig. \ref{fig:plaxton_mc}). In sum, the expression for routability can be 
succinctly given as follows: $r = \frac{(2-q)^d-1}{(1-q)2^d-1}$
\subsubsection{XOR}
As reviewed in section \ref{sec:overview}, connecting to a neighbor
at an XOR distance of $[2^{d-i},2^{d-i+1}]$ is equivalent to choosing a 
neighbor by matching the first (i-1) bits of one's identifier, flipping the $i$th bit, 
and choose random bits for the rest of the bits.  Note that this is equivalent to 
how neighbors are chosen in the Plaxton-tree routing geometry. As a result, 
the $n(h)$ expression is given as: $n(h) = \binom{n}{h}$ just as in the tree case.

Now, let's examine how the Markov chain model (Fig. \ref{fig:xor_mc}) is obtained: 
in this scenario, a message is to be routed to a destination $h$ phases away; starting at state 
$S_0$, state $S_1$ is reached if 
the optimal neighbor correcting the leftmost bit exists, which happens with probability $1-q$ 
($S_i$ denotes the state that corresponds to the $i$th advanced phase). 
However, if all $h$ neighbors have failed (i.e. with probability $q^h$), the failure state $F$ 
is entered.  Otherwise, the routing process can correct one of the lower order bits, which 
happens with probability $q(1-q^{h-1})$. Note that there is a maximum number of $h-1$ lower order
bits that can be corrected in the first phase.  All other transition probabilities can be obtained similarly. 
By inspecting the Markov chain model, 
we note that $G(S_0,S_1) = 1-Q_{xor}(h)$, $G(S_1,S_2) = 1-Q_{xor}(h-1)$, 
and so forth, where the function $Q_{xor}(m)$ is defined as follows: 
\begin{eqnarray}
Q_{xor}(m)&=&q^m + \sum_{k=1}^{m-1}q^m[\displaystyle\prod^{m-1}_{j=m-k} (1-q^j)] \label{eq:xor_q} \\
            &\approx&q^{m}(m+\frac{q}{1-q}(q^{m-1}(m-1)-\frac{1-q^{m+1}}{1-q})) \nonumber 
\end{eqnarray}
The approximation is obtained by invoking the following: $1-x \approx e^{-x}$ for x small.

\begin{figure*}[htp]
\begin{subfigmatrix}{2}
\subfigure[\small System Size N=$2^{100}$]{
\label{fig:asymptotic}
\centering
\includegraphics[width=3.35in]{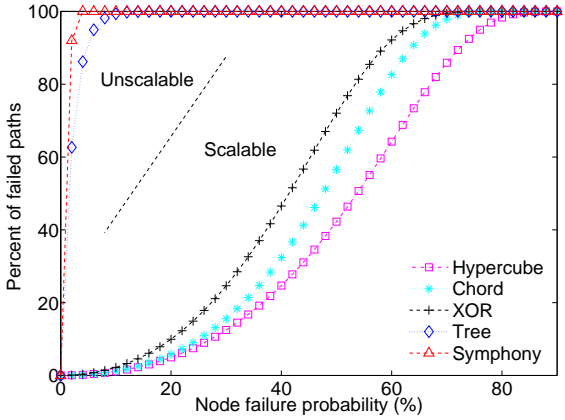}
}
\subfigure[\small Failure Probability q=0.1]{
\label{fig:scale_curves}
\centering
\includegraphics[width=3.35in]{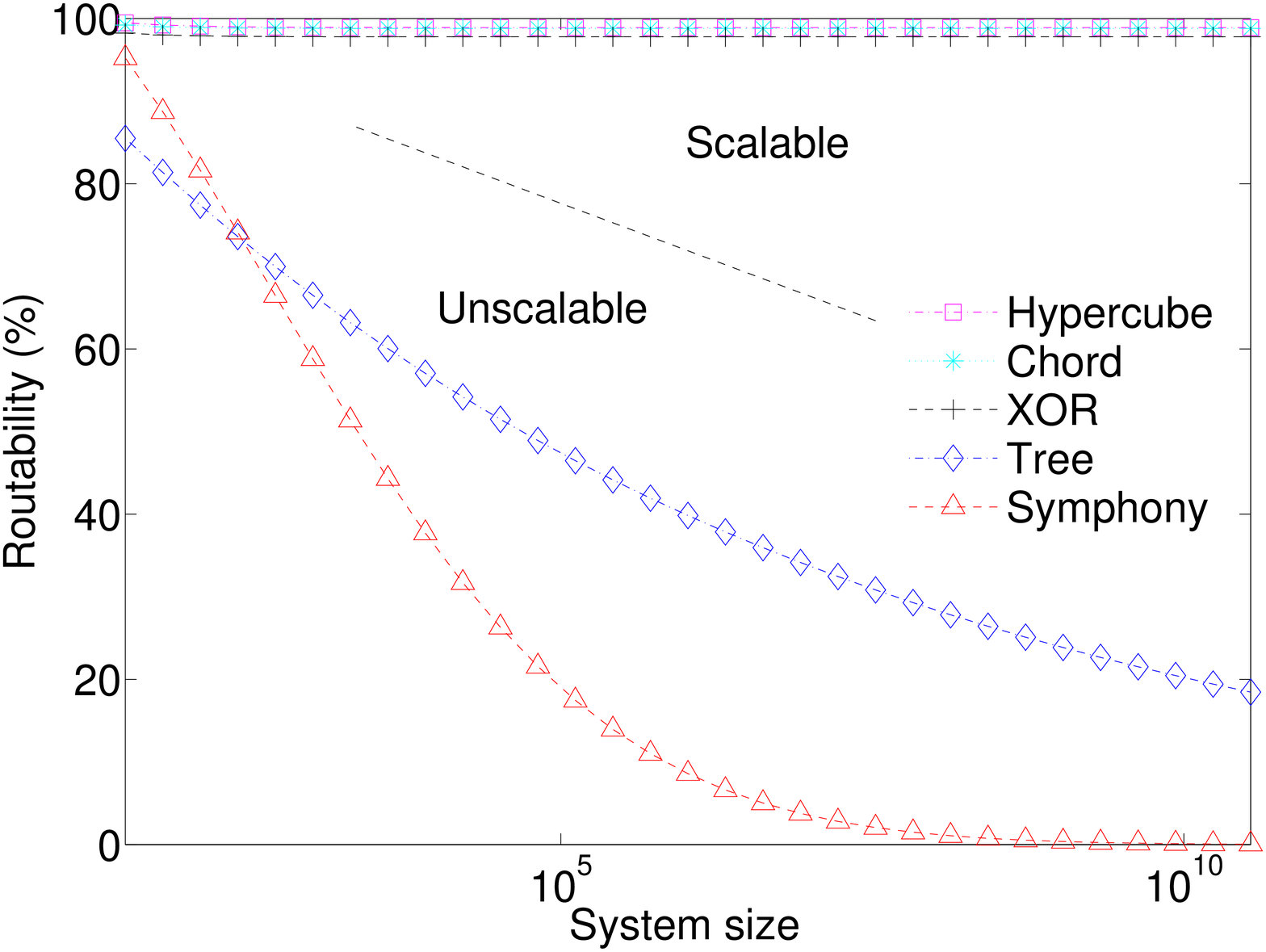}
}
\end{subfigmatrix}
\caption{\footnotesize (a) Asymptotic limit: This figure plots the percentage of failed paths 
(i.e. 1-routability) for varying
node failure probability in the asymptotic limit.  The curves are obtained by evaluating the 
analytical expressions at N=$2^{100}$.  Note that the curves for tree and symphony are very 
close to a step function, which is consistent with our analysis.  In addition, the curves for 
the other three geometries are very close to the case for N=$2^{16}$. (b) Routability vs N: 
This plot shows the routability of the routing geometries for varying system size and a constant 
failure probability (q=0.1).  This figure clearly demonstrates the lack of scalability of the tree and 
Symphony routing geometries. As the system scales, the routability of both the tree and Symphony routing 
systems monotonically degrades toward zero. In contrast, the other three geometries remain highly 
routable in the face of failure even as the systems scale to billions of nodes. (In both plots, 
we set the number of near neighbors and the number of shortcuts equal to one for Symphony.)  }
\end{figure*}

\subsubsection{Ring}
In ring routing as implemented in Chord, when a node takes a suboptimal hop in the routing process, 
the progress made by taking this suboptimal hop is preserved in later hops.  For example, 
consider the scenario that a message is to be routed to a node at a numeric distance that is $O(N)$
(i.e. the message is to be routed one full circle around the ring), and the fingers are connected 
to nodes that are half way across the ring, one quarter across the ring, etc.  
For the message's first hop, it takes a suboptimal hop which takes the message only one quarter across 
the ring, because the finger that would have taken the message half way across the ring has failed.  
Then, for the message's second hop, none of the finger connections has failed.  Thus, the message 
takes an optimal hop which takes the message half way across the ring.  Therefore, after two hops, 
the message is now three quarters of the way across the ring.  Note that the progress made in the 
first suboptimal hop is this scenario is later preserved by a subsequent hop. 
 
This property that suboptimal hops in ring routing contribute non-trivially to the routing process is 
not accounted for in the the Markov chain model as illustrated in Fig. \ref{fig:chord_mc}.  The 
reason is that accounting for progress made by suboptimal hops would lead to an exponential blowup 
in the number of terms that we need to keep track of for computing $p(h,q)$.  This simplified 
Markov chain model essentially makes the assumption that progress made 
by suboptimal hops do not contribute to the routing process.  Therefore, the analytical expression 
for $p(h,q)$ using this model provides a \emph{lower bound}.  

The Markov chain model for ring routing \ref{fig:chord_mc} is very similar to the one 
for XOR routing (Fig. \ref{fig:xor_mc}).  However, fundamental differences exist: first, when a suboptimal hop is taken 
in Chord, the number of next hop choices does not decrease.  For example, in the first phase, 
there are $h$ choices for the next hop, thus the transition probabilities from the states 
in the first phase to the failure state are given by $q^h$. In contrast, the corresponding 
transition probabilities in Fig. \ref{fig:xor_mc} are given by $q^h$, $q^{h-1}$, and so forth.
In addition, the maximum number of suboptimal hops in Chord is given by $2^{h-1}$, $2^{h-2}$ 
and so forth, while the corresponding transition probabilities in Fig. \ref{fig:xor_mc} are 
given by $h$, $h-1$, and so forth. This difference is due to the fact that in XOR routing, 
routing fails if all the lower order bits are resolved and the leftmost bit is not yet resolved.
However, Chord does not have such restriction.The results for the ring routing geometry 
is derived by inspecting Fig. \ref{fig:chord_mc}:
\begin{align*}
 Q_{ring}(m) &= q^m\sum_{k=0}^{2^{m-1}-1}[q(1-q^{m-1})]^k \\
             &= q^{m}\frac{1-[q(1-q^{m-1})]^{2^{m-1}}}{1-q(1-q^{m-1})}  
\end{align*}
In addition, one can easily see by inspection that the $n(h)$ expression for the ring geometry 
is given by: $n(h) = 2^{h-1}$.  

\subsubsection{Symphony}
\begin{figure*}[htp]
\begin{subfigmatrix}{2}
\subfigure[\small Markov chain model for ring routing]{
\label{fig:chord_mc}
\centering
\includegraphics[width=3.35in,height=3.5in]{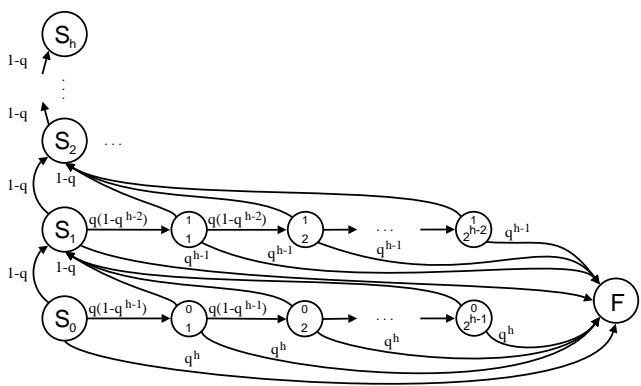}
}
\subfigure[\small Markov chain model for Symphony routing]{
\label{fig:symphony_mc}
\centering
\includegraphics[width=3.35in,height=3.5in]{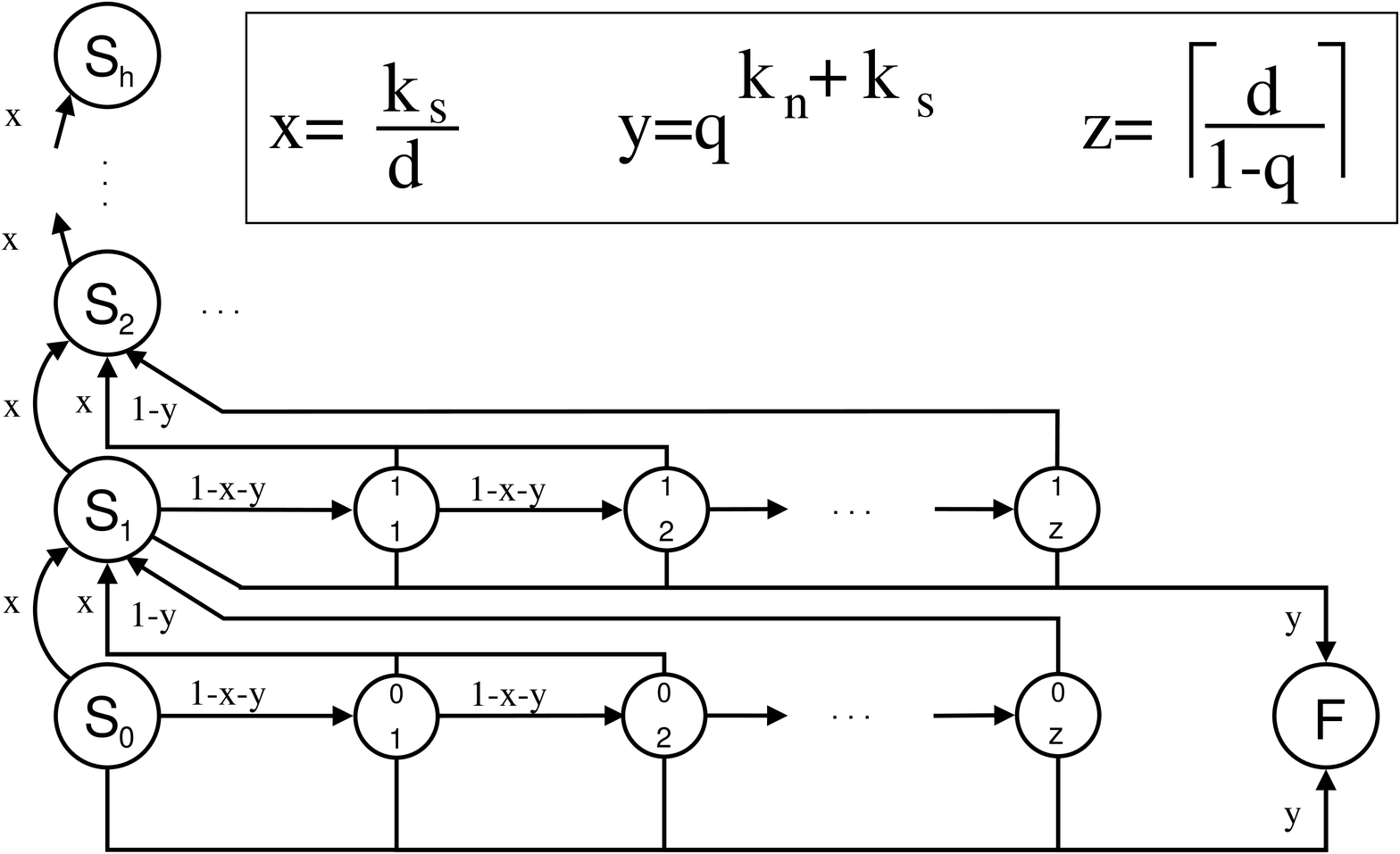}
}
\end{subfigmatrix}
\caption{\small The above two diagrams illustrate the Markov chain model for ring and Symphony routing 
geometries.}
\end{figure*}

Symphony's Markov chain model (Fig. \ref{fig:symphony_mc}) is fundamentally 
different from the ones for XOR routing (Fig. \ref{fig:xor_mc}) and ring routing 
(Fig. \ref{fig:chord_mc}). Starting at $S_0$, one phase is advanced if any of the node's 
shortcuts connects to the desired phase, which happens with probability $\frac{k_s}{d}$ where
$k_s$ denotes the number of shortcuts.  Alternatively, the routing fails if all of the node's
near neighbor and shortcut connections fail, which happens with probability $q^{k_n+k_s}$.
The third possibility is taking a suboptimal hop, which happens with probability 
$1-\frac{k_s}{d}-q^{k_n+k_s}$.  All other transition probabilities in the Markov chain 
can be similarly derived.  Note that we approximate the maximum number of suboptimal 
hops by $\lceil \frac{d}{1-q} \rceil$. 

For the Symphony routing geometry, we note that the expression for the $Q$'s is constant for all phases.  
The results are similarly derived as the other systems by inspecting Fig. \ref{fig:symphony_mc}:
\begin{align}
 Q_{sym} &= q^{k_n+k_s}\sum^{\lceil\frac{d}{1-q}\rceil}_{j=0}(1-\frac{k_s}{d}-q^{k_n+k_s})^j \nonumber \\
         &\approx q^{k_n+k_s}( \frac{1-(1-\frac{k_s}{d}-q^{k_n+k_s})^{\frac{d}{1-q}+1}}{1-(1-\frac{k_s}{d}-q^{k_n+k_s})})
\label{eq:symphony_q}
\end{align}
The symbols $k_n$ and $k_s$ denote the number of near neighbors and shortcuts respectively.  
Similarly to ring routing, the $n(h)$ expression for the Symphony routing algorithm
is given by: $n(h) = 2^{h-1}$.  

\section{Scalability of DHT Routing Protocols under Random Failure}\label{sec:scalability}

For a DHT routing system to be scalable, its routability must converge
to a non-zero value as the system size goes to infinity (Definition \ref{def:scalability}). Alternatively, we examine 
the asymptotic behavior of $p(h,q)$ with $h$ set to the average routing distance in the system 
(i.e. $h=O(\log N)$ or $O(\log^2 N)$ for Symphony).  Using Eq. \ref{eq:routability_eq}, it is 
simple to show that the equivalent condition for scalability is as follows:
\begin{equation}
 \lim_{N \rightarrow \infty} p(h,q) = \lim_{h \rightarrow \infty} p(h,q) > 0  \ \mbox{for} \ 0 < q < 1-p_c
\label{scalable_cond}
\end{equation}
Otherwise, the routing system is unscalable.  In other words, the equivalent condition 
for system scalability states 
that as the number of routing hops to reach a destination node in the system approaches 
infinity, the probability of successfully routing to the destination node must not drop 
to zero for a non-zero node failure probability in the system.  

As discussed in section \ref{ssec:summary_other}, all of the DHT systems under study have 
the property that the probability of successfully traveling $h$ hops or phases from the 
root node is given by the following form:
\begin{equation}
 p(h,q) = \prod^h_{m=1} (1-Q(m))
\label{scalable_eq}
\end{equation}
where $Q(m)$ can be thought of as the probability of failure at the $m$th phase of the
routing process.  

\begin{qTheorem}
{(\emph{From Knopp} \cite{Knopp:inf_series})}
\label{series_thm}
If, for every $n$, $0 \le a_n < 1$, then the product $\prod (1-a_n)$ tends to a limit greater 
than 0 if, and only if, $\sum a_n$ converges.
\end{qTheorem}

Theorem \ref{series_thm} allows us to conveniently convert our problem of
determining the convergence of an infinite product to a simpler infinite sum.
Thus, $p(h,q)$ is convergent if and only if $\sum Q(m)$ converges. 
\subsection{Tree}
The case for the tree routing geometry can be trivially shown to be \emph{unscalable}:
\begin{equation}
\lim_{h \rightarrow \infty}(1-q)^h = 0 \ \ \mbox{for}\ \mbox{any}\ q > 0
\end{equation}
\subsection{Hypercube}
For hypercube routing, $p(h,q)$ is given by $p(h,q) = \displaystyle\prod^{h}_{m=1}(1-q^m)$ 
(Eq. \ref{eq:hypercube_eq}).  By invoking Theorem \ref{series_thm}, it is trivial to see that 
$\sum q^m$ converges for $0 < q < 1-p_c$.  Thus, the hypercube routing geometry is \emph{scalable}.  
\subsection{XOR}
In XOR routing, the $Q(m)$ expression given by Eq. \ref{eq:xor_q}.  It is simple to show that the $Q(m)$ series 
involves only $q^m$ and $mq^m$ terms.  Thus, $\sum Q(m)$ is convergent and the XOR routing 
scheme is \emph{scalable}.   
\subsection{Ring}
We will demonstrate that the ring routing geometry is also scalable by showing that the 
XOR results derived above is a lower bound for the ring geometry.
We compare the Markov chain models for the ring geometry  and the XOR geometry 
(Fig. \ref{fig:chord_mc} and Fig. \ref{fig:xor_mc}).  We note that the transition 
probabilities for the suboptimal hops in ring are strictly greater than the corresponding probabilities 
for XOR.  For example, in Fig. \ref{fig:chord_mc}, note that the transition probabilities 
for $S_0 \rightarrow (0,1)$, $(0,1) \rightarrow (0,2)$ and so forth 
are given by $q(1-q^{h-1})$.  These probabilities are strictly greater than the corresponding 
transition probabilities in Fig. \ref{fig:xor_mc}.  Thus, by comparing these two Markov chain 
models, it is simple to show that the $p(h,q)$ expression for the ring routing geometry is 
strictly greater than the $p(h,q)$ expression for XOR routing.  Thus, the ring routing 
geometry is also \emph{scalable}.
\subsection{Symphony}
In Symphony routing, the $Q(m)$ expression given by Eq. \ref{eq:symphony_q}.  Note that 
the $Q(m)$ expression is given by a 
constant term. Therefore, $\sum Q(m)$ is divergent and the Symphony routing scheme is \emph{unscalable}.  
\\ \\
Please refer to Fig. \ref{fig:asymptotic} and \ref{fig:scale_curves} for plots of the above 
scalability results.

\section{Concluding Remarks}\label{sec:conclusion}
In this work, we present the reachable component method (RCM) which is an 
analytical framework for characterizing DHT system performance
under random failures. The method's efficacy is demonstrated through an
analysis of five important existing DHT systems and the good agreement of the RCM
predictions for each system with simulation results from the literature. 
Researchers involved in P2P system design and implementation can use the method 
to assess the performance of proposed architectures and to choose robust routing 
algorithms for application development.  In addition, although the analysis
presented in this work assumes fully-populated identifier spaces, 
analytical results for real world DHTs with non-fully-populated identifier 
spaces can be similarly derived.  Detail investigation in this area will be 
left for future work.  

One of the most interesting implications of this analysis is that
in the large-network limit, some DHT routing systems are incapable
of routing to a constant fraction of the network if there is any 
non-zero probability of random node failure. These 
DHT algorithms are therefore considered to be \emph{unscalable}.  Other
algorithms are more robust to random node failures, allowing each node 
to route to a constant fraction of the network even as the system size goes to
infinity. These systems are considered to be \emph{scalable}.
Now that real DHT implementations have on the order of millions of highly transient nodes,
it is increasingly important to characterize how the size and failure conditions of a
DHT will affect its routing performance. 

\section{Acknowledgments}
We would like to thank Krishna Gummadi for furnishing the simulation results for 
DHT systems. We wish to thank Nikolaos Kontorinis for his feedback and suggestions.
This work was in part supported by the NSF grants ITR:ECF0300635 and BIC:EMT0524843.

\bibliographystyle{latex8}  
\bibliography{routing_rcm}% Produces the bibliography via BibTeX.
 
\end{document}